\documentclass[fleqn,twoside]{article}
\usepackage{espcrc2}
\usepackage{amsmath}
\usepackage{amsfonts}
\usepackage{amssymb}
\usepackage{wasysym}
\usepackage{graphicx}
\usepackage[figuresright]{rotating}

\title{UV AND IR behaviour for QFT and LCQFT with fields as Operator Valued Distributions: Epstein and Glaser revisited}

\author{Pierre Grang\'e\address[MCSD]{Laboratoire de Physique Th\'eorique et Astroparticules,
UMR-CNRS $5207$,\\  Universit\'e de Montpellier II,$34060$ Montpellier-Cedex,
France.}
 and  Ernst Werner\address{ Institut f\"{u}r Theoretische Physik,\\ Universit\"{a}t
Regensburg,\\D$-93040$,Regensburg, Germany.}}
       
\begin{document}

\begin{abstract}
Following Epstein-Glaser's work we show how a QFT
formulation based on operator valued distributions (OPVD) with adequate test
functions treats original singularities of propagators on the diagonal in
a mathematically rigourous way.Thereby  UV and/or IR divergences are avoided
at any stage, only a finite renormalization finally occurs at a point related to the
arbitrary scale present in the test functions.Some well known UV
cases are examplified.The power of the IR treatment
is shown for the free massive scalar field theory developed in the (conventionally hopeless) mass 
perturbation expansion.It is argued  that the approach should
prove most useful for non pertubative methods where the usual determination of 
counterterms is elusive.
\vspace{1pc}
\end{abstract}

\maketitle

\section{INTRODUCTION}
Since the developments of Quantum Fields Theory (QFT) of the early sixties \cite{Bolo,Schw} it is almost
a trivial statement to mention that fields are operator valued distributions (OPVD). However it is soon
 forgotten as one starts multiplying  fields at the same space-time point as if they were regular functions.The
results are mathematically undefined expressions,divergences and the chase for counter terms
etc...However the use of distributions cannot be separated from the class of test functions necessary
to give a well defined functional integral and to validate all the usual operations
(translation,derivations,Fourier transforms,etc...) in the distributional context.In QFT there are many reasons to
use test functions.In Light Cone QFT (LCQFT) one reason is particularly compelling: it has to do with
the consistency of the canonical quantization scheme itself.It is best seen for the massive scalar
 field.The LC-Laplace operator is linear in the light-cone time and initial data on two charateristics are
necessary. Canonical quantization in terms of initial field values in the light cone time is possible
provided \cite{THEW} $\lim_{p^{+} \rightarrow 0}\frac{\chi(p^{+})}{p^{+}}=0$,where $\chi(p^{+})$ is the
field amplitude at $p^{+}=p^{0}+p^{1}$. With fields as OPVD this relation becomes
 $\lim_{p^{+} \rightarrow 0 }\frac{f(p^{+})}{p^{+}}$, which is surely satisfied for the class of test functions 
 $f(p^{+})$ in
accordance with the nature of distributions in use.It is only recently that we realize the
filiation of our LCQFT approach with test functions\cite{GSW,GUW} to the early work of Epstein and Glaser
\cite{EG73}.Following some recent developments in mathematical physics \cite{ULL,Mphys,GBL03} it is our aim here 
to show how Epstein and Glaser's treatment of mathematically undefined expressions carries over with 
important simplications when using test functions which are partitions of unity,as advocated in earlier 
publications and LC-workshops.

\section{FIELDS AS OPVD}

To introduce fields  as OPVD one may consider,without loss of generality,
the free massive  scalar field in D-dimension.The Klein-Gordon (KG) equation , $(\Box_x + m^2)\varphi(x) = 0$, 
writes, after a Fourier transform, $(p^2-m^2)\tilde{\varphi}(p)=0$. The solution is a distribution 
$\tilde{\varphi}(p)=\delta^{(D)}(p^2-m^2) \chi(p)$, with $\chi(p)$ arbitrary. The solution of the KG-equation
 is therefore also a distribution, {\it{ie}} an 
OPVD, which defines a functional with respect to a test function $\rho(x)$, which is $C^{\infty}$ with compact
support,
\begin{equation}
\Phi(\rho) \equiv <\varphi,\rho> = \int d^{(D)}y\varphi(y)\rho(y).
\end{equation}
Here $\Phi(\rho)$ is an operator-valued functional with the possible interpretation of a more general functional 
$\Phi(x,\rho)$ evaluated at $x=0$.Indeed the translated functional is a well defined object \cite{Sch63} such that 
\begin{eqnarray}
T_{x}\Phi(\rho) &=& <T_{x}\varphi,\rho> = <\varphi,T_{-x}\rho> \nonumber\\
&=& \int d^{(D)}y \varphi(y) \rho(x-y)
\end{eqnarray}
Now the test function $\rho(x-y)$ has a well defined Fourier decomposition
\begin{equation}
\rho(x-y) = \int{d^{(D)}q \over (2 \pi)^{D}}\exp^{iq(x-y)}f(q)
\end{equation}
It follows that
\begin{equation}
T_{x}\Phi(\rho) = \int{d^{(D)}p \over (2 \pi)^{D}} e^{-ipx}\delta(p^2-m^2)\chi(p)f(p).
\end{equation}
Due to the properties of $\rho$ $, T_{x}\Phi(\rho)$ obeys the KG-equation and is taken as the physical field with quantized
form
\begin{equation}
\varphi_{1}(x) = \int{d^{(D-1)}p \over (2 \pi)^{D-1}}{f(p,\omega_{p}) \over (2 \omega_{p})}[a^{+}_{p} e^{ipx} + a_{p} e^{-ipx}].
\end{equation}
$f(p,\omega_p)$ acts as regulator \cite{GUW,ULL} with very specific properties \footnote{f(p) is also $C^{\infty}$ with fast decrease
in the sense of L. Schwartz \cite{Sch63}}. This expression for $\varphi_{1}(x)$ is particurlarly useful on the LC because
 the Haag-series can be used \cite{GSW}  and is well defined in terms of the product of $\varphi_{1}(x_i)$. In Euclidean metric,relevant for the sequel,there is no on-shell condition and $\varphi_{1}(x)$ stays a $D$-dimensional Fourier transform with $f(p) \rightarrow f(p^2)$.
It may appear that
 there would be as many QFT as eligible test functions.However the paracompactness 
 property of the Euclidean manifold permits using test functions which are partition of unity \cite{KOB} and the resulting operator-valued
 functional is indepedent of the way this partition of unity is constructed \cite{Sch63}.Then $f(p^2)$ is $1$ except in the boundary
  regions where it is $C^{\infty}$ and goes to zero with all its derivatives.The dimensionless nature of $f(p^2)$ implies the presence
  of an arbitrary scale directly related to the renormalization group analysis of the physical observables.

\subsection{ The Euclidean Epstein and Glaser approach in a nutshell. The magic of Lagrange's formula for the Taylor remainder.}

Epstein and Glaser chose working in Minkowskian metric.There are two main aspects to their original work.The first one relates 
to the implementation of causality in their building up of the $S$-matrix leading to the generic name of "Causal Pertubation Theory"
\cite{Scharf}.The second one deals with the treatment of specific divergences encountered when multypling fields at the same 
space-time point.Since this is our main concern here, and to avoid causal issues, we shall work in Euclidean metric where the use of
test functions as partition of unity is well founded \cite{KOB}.\\
In standard massive scalar field theory the propagator $\Delta(x-y)$ is a well known example of a divergence occuring 
when $x=y$.Using the Euclidean counterpart of the definition of $Eq.(4)$ for the fields,$\Delta(x-y)$ reads
\begin{equation}
\Delta(x-y)=\int \frac{d^{D}p}{(2\pi)^D}\frac{e^
{[-i p.(x-y)]} f^2(p^2)}{(p^2+m^2)} .
\end{equation}
At $D=2..4$  and for $x \neq y $   $\Delta(x-y)$  is finite and  $f^2(p^2)$  may be taken to $1$  everywhere.One of our aims is to 
understand the role of the partition of unity in the extension of  $\Delta(x-y)$  to the diagonal.\\
We turn now to Epstein and Glaser analysis of singular distributions.Consider a distribution $T(X)$ singular 
at the origin of ${\mathbb R}^{d}$. Then $T(X) \in ({\mathbb S}^{\prime}({\mathbb
 R}^{d})\backslash\{0\})$.Its singular order $k$ is defined as
\begin{equation}
 k=inf \{s:{\displaystyle \lim_{\lambda \rightarrow 0}}\lambda ^{s} T(\lambda X) = 0\} - d
\end{equation} 
 The aim is to extend $T(X)$  to the whole domain ${\mathbb S}^{\prime}({\mathbb
 R}^{d})$.If $f(X) \in  {\mathbb S}({\mathbb R}^{d})$ is a general test function Epstein and Glaser 
 perform a Taylor series sugery by throwing away a weigthed $k-$jet of $f(X)$ at the origin and deal 
 with the Taylor remainder $R^{k}_{0}f$.The following operation \cite{GBL03} 
\begin{equation}
{\mathbb P}^{w}f(X)=(1-w(X))R^{k-1}_{0}f(X)+w(X)R^{k}_{0}f(X),\nonumber
\end{equation} 
defines now a new {\it bona fide} test fuction in $ {\mathbb S}({\mathbb R}^{d})$ which regulate
 at the origin the singular behaviour of $T(X)$.Here $w(X)$ is Epstein-Glaser's weight function such that
$w(0)=1,w^{(\alpha)}(0)=0,0 <\mid \alpha \mid\leq k$ The extension $\widetilde{T}(X)$ of $T(X)$ is defined from the
relations
\begin{equation}
<\widetilde{T},f> =<T, {\mathbb P}^{w}f>=\!\!\!\int \!\!d^{d}\!X  T(X)  {\mathbb P}^{w}f(X).
\end{equation}
The important observation \cite{Mphys,GBL03} is that Lagrange's expression for the Taylor remainder $R^{k}_{0}f$ 
permits the necessary partial integrations in the above integral to extract $\widetilde{T}(X)$.They are given
respectively by
\begin{eqnarray}
R^{k}_{0}f(X)\!\!\!\!&=&\!\!\!\!(k+1){\displaystyle \sum_{\mid \beta \mid=k+1}}\big[
\frac{X^{\beta}}{\beta !}\int^{1}_{0} dt (1-t)^{k} \nonumber \\
\!\!\!& &\!\!\!\partial^{\beta}_{(tX)}f(tX)\big],
\end{eqnarray}
and
\begin{eqnarray} 
\widetilde{T}(X)\!\!\!\!&=&\!\!\!\!(-)^{k} k {\displaystyle \sum_{\mid \alpha \mid=k}}\partial^{\alpha}\big[
\frac{X^{\alpha}}{\alpha !}\int^{1}_{0} dt \frac{(1-t)^{k-1}}{t^{k+d}}T(\frac{X}{t})\nonumber \\
\!\!\!\!& &\!\!\!\!(1-w(\frac{X}{t})\big] 
+ (-)^{k+1} (k+1) {\displaystyle \sum_{\mid \alpha \mid=k+1}}\partial^{\alpha} \nonumber \\
\!\!\!\!& &\!\!\!\!\big[
\frac{X^{\alpha}}{\alpha !}\int^{1}_{0} dt \frac{(1-t)^{k}}{t^{k+d+1}}T(\frac{X}{t})w(\frac{X}{t})\big].
\end{eqnarray}

\subsection{ Partition of unity: example and properties.}

With a test function $f(X)$ reduced to a partition of unity on a given domain of the Euclidean
manifold important simplifications in the above formalism occur.Here the domain is the ball $B_{1+h}(\| X \|)$ around  
$\| X\|=0$  of radius $1+h$ and $f^{(\alpha)}(0)=f^{(\alpha)}(1+h)=0$ ,  $\forall \alpha \geq 0$. In the boundary regions 
the test function is strictly equal to its Taylor remainder of any finite order $k.$ $\forall k \geq 0$ at 
$\| X \| \approx 1+h$ it holds that
\begin{eqnarray}
f(X) \equiv f^{>}(X) \!\!\!\!&\equiv&\!\!\!\! -(k+1)\!\!\!\!\!{\displaystyle \sum_{\mid \beta \mid=k+1}}\!\!\!\big[
\frac{X^{\beta}}{\beta !}\int^{\infty}_{1} \!\!dt (1-t)^{k} \nonumber \\
\!\!\!\!& &\!\!\!\!\partial^{\beta}_{(tX)}f(tX)\big], 
\end{eqnarray}
and at $\| X \| \approx 0$ 
\begin{eqnarray}
f(X) \equiv f^{<}(X) \!\!\!\!&\equiv&\!\!\!\! (k+1)\!\!\!\!{\displaystyle \sum_{\mid \beta \mid=k+1}}\!\!\!\big[
\frac{X^{\beta}}{\beta !}\int^{1}_{0} dt (1-t)^{k} \nonumber \\
\!\!\!\!& &\!\!\!\!\partial^{\beta}_{(tX)}f(tX)\big]. 
\end{eqnarray}
Hence $f^{\gtrless}(X)$ give respectively the ultraviolet and infrared extensions $\widetilde{T}^{\gtrless}(X)$ of $T(X)$.
$f^{>}(X)$ is such that $f^{>}(X)\!\!\!=\!\!\!\{ 1\ \mathrm{for~} \| X \| \!\!\!\leq 1; \mathrm{\chi(\|X\|,h)} \ \mathrm{for~}
 1 < \| X \| \leq 1+h;$  $ 0  \  \mathrm{for~} \| X \|  > 1+h \}$. Because of the indepedence of the procedure on the specific
 construction of this partition of unity \cite{Sch63} its precise expression is not necessary. To fix ideas a possible choice
of $\mathrm{\chi(\|X\|,h)}$ is 
\begin{eqnarray}
\chi(\|X\|,h)\!\!\!\!&=&\!\!\!\!{\mathbb N}_{h} \int^{h}_{\| X \|-1} e^{[\frac{h^2}{v(v-h)}]} dv ,
\end{eqnarray}
where the requirement $\chi(1,h)=1$ fixes the normalisation ${\mathbb N}_{h}$.
This function effectively builds up unity because of the property that $\mathrm{for~}\ (1-h) < \| X \| \leq 1,
\mathrm{\chi(2-\|X\|,h)}+\mathrm{\chi(\|X\|+h,h)} = 1$. Here $h$ is a parameter which may depend on $\| X \|$.The 
consequences are then\\
\vspace{0.2cm}
$-i)$ $\exists$    $\|X\|_{max}$   such that   $\|X\|_{max}=1+h(\|X\|_{max})\equiv 
\mu^2 \|X\|_{max} g(\|X\|_{max})$ $\Longrightarrow$ $g(\|X\|_{max})=\frac{1}{\mu^2}$,\\
\vspace{0.2cm}
\hspace{0.1cm}  $-ii)$   $h >0$ $\Longrightarrow$ $\mu^2\|X\| g(\|X\|) > 1 $ $\forall$  $\|X\|$ $\in$  $[1,\|X\|_{max}]$ 
$\Longrightarrow$  $g(1) > g(\|X\|_{max})$, \\
\vspace{0.2cm}
\hspace{0.1cm}  $-iii)$ from $f^{>}(Xt)$ present in Lagrange's formula
one has  $t < \frac{1+h(\|X\|)}{\|X\|}= \mu^2 g(\|X\|)$. \\
In the definition of $h(\|X\|)$ a dimensionless scale factor $\mu^2$ has been extracted from $g(\|X\|)$ for the
purpose of later discussion.

\section{ ULTRAVIOLET EXTENSION OF $T(X)$. }

From the expression of $f^{>}$ the UV-extension $\widetilde{T}^{>}(X)$ of $T(X)$ is such that
\begin{eqnarray}
<T,f^{>}>\!\!\!\!&=&\!\!\!\!\int \!\!d^{d}X  T(X)
         \big{\{}\!\!-(k+1)\!\!\!\!\!{\displaystyle \sum_{\mid \beta \mid=k+1}}\!\!\!\big[
\frac{X^{\beta}}{\beta !} \nonumber\\
\!\!\!\!& &\!\!\!\!\int^{\!\!\ ^{\mu^2 g(X)}}_{1} dt \frac{(1-t)^{k}}{t^{(k+1)}}
\partial^{\beta}_{X}f^{>}(tX)\big]\big{\}} \nonumber \\
 \!\!&=&\!\!<\widetilde{T}^{>},1>, 
\end{eqnarray}
where in the last line the partial integrations in $X$ have been performed giving
\begin{eqnarray}
\widetilde{T}^{>}(X)\!\!\!\!&=&\!\!\!\!(-)^{k}(k+1){\displaystyle \sum_{\mid \beta \mid=k+1}} \partial^{\beta}_{X}
\big[\frac{X^{\beta}}{\beta !} T(X) \nonumber \\
\!\!\!\!& &\!\!\!\!\int^{\!\!\ ^{\mu^2 g(X)}}_{1}\hspace{-0.5cm}dt
\frac{(1-t)^{k}}{t^{(k+1)}}\big].
\end{eqnarray}
An immediate application of this relation is for the scalar propagator at $x=y$. Setting $\|X\| \equiv  X$ from now
on, we have $X=\frac{p^2}{\Lambda^2}$,$T(X)=\frac{1}{X \Lambda^2+m^2}$ and at $D=2$ the dimension in the $X$
variable is $d=1$ and $k=0$. Then
\begin{eqnarray}
\widetilde{\Big[\frac{1}{(p^2+m^2)}}\Big]_{\mu,D=2}\!\!\!\!\!&=&\!\!\!\!\partial_{X}\big[\frac{X}{(X \Lambda^{2}+m^{2})}
\int^{\!\!\ ^{\mu^2  g(X)}}_{1}\!\!\frac{dt}{t}\big] \nonumber \\
\!\!\!\!\!&=&\!\!\!\!\frac{m^2 \log[\mu^2  g(X)]}{(X \Lambda^{2}+m^{2})^2} \nonumber \\
\!\!\!\!\!& &\!\!\!\!+\frac{X  g^{\prime}(X)}{(X \Lambda^{2}+m^{2})g(X)}.
\end{eqnarray}
It is clear that the choice $g(x)=x^{(\alpha-1)}$ (up to a multiplicative arbitrary constant already taken into account as $\mu^2$) {\it ie} $h(x)=\mu^2 x^{\alpha}-1$  with $0< \alpha <1$ is consistent with
the contruction of $\chi(X,h)$. It implies also $g(1)=1 > g(X_{max})=\frac{1}{\mu^2}$  {\it ie} $\mu^2>1$ and 
$X_{max}=(\mu^2)^{(\frac{1}{(1-\alpha)})}$.In the limit $\alpha \rightarrow  1$ $\frac{X g^{\prime}(X)}{g(X)}=0$ thereby eliminating the
last term of Eq.(16) and extending the upper integration limit in $X$ to infinity. In this limit the propagator at $x=y$ is then given
by
\begin{eqnarray}
\Delta(0)\!\!\!\!\!&=&\!\!\!\!\!\!\int \frac{d^{2}p}{(2\pi)^2}\frac
{f^2(p^2)}{(p^2+m^2)}=\int \frac{d^{2}p}{(2\pi)^2}\frac
{m^2 \log(\mu^2)}{(p^2+m^2)^2} \nonumber \\
\!\!\!\!\!&=&\!\!\!\! \frac{1}{(4\pi)}\log(\mu^2), 
\end{eqnarray}
which is RG-invaraint with respect to the scale parameter $\mu$.At Euclidean dimension $D=4$ one has $d=2,k=1$.Then
\begin{eqnarray}
\widetilde{\Big[\frac{1}{(p^2+m^2)}}\Big]_{\mu,D=4}\!\!\!\!\!&=&\!\!\!\!\!\!\lim_{\alpha \rightarrow 1} -\partial^{(2)}_{X}
\big[\frac{X^2}{(X \Lambda^{2}+m^{2})} \nonumber \\
\!\!\!\!\!& &\!\!\!\!\!\!\int^{\!\!\ ^{\mu^2 g(X)}}_{1}\!\! dt \frac{(1-t)}{t^2}\big]  \\
\!\!\!\!\!&=&\!\!\!\!\!\!\frac{2 m^4}{\mu^2}\frac{[1-\mu^2+\mu^2 \log(\mu^2)]}{(X \Lambda^{2}+m^{2})^3}. \nonumber
\end{eqnarray}
Integrating over $X$ ({\it viz.} $p^2$) gives the familiar result
\begin{eqnarray}
\Delta(0)\!\!\!\!\!&=&\!\!\!\!\!\!\int \frac{d^{4}p}{(2\pi)^4}\frac
{2 m^4}{\mu^2}\frac{[1-\mu^2+\mu^2 \log(\mu^2)]}{(p^2+m^2)^3} \nonumber \\
\!\!\!\!\!&=&\!\!\!\! \frac{1}{(8\pi^2)}\frac
{m^2}{2\mu^2}[1-\mu^2+\mu^2 \log(\mu^2)]. 
\end{eqnarray}
There is an alternative form of $\widetilde{T}^{>}(X)$ which is quite instructive. It is obtained trough the change of variable
$Xt \rightarrow Y$ in Eq.(14).It gives 
\begin{eqnarray}
\widetilde{T}^{>}(X)\!\!\!\!\!&=&\!\!\!\!\!(-)^{k}(k+1){\displaystyle \sum_{\mid \beta \mid=k+1}} \partial^{\beta}_{X}
\big[\frac{X^{\beta}}{\beta !} \nonumber \\
\!\!\!\!\!& &\!\!\!\!\!\int^{\!\!\ ^{\mu^2}}_{1}dt\frac{(1-t)^{k}}{t^{(k+d+1)}} T(X/t)\big].
\end{eqnarray}
The scalar propagator at $D=2$ and $x=y$ now becomes 
\begin{eqnarray}
\widetilde{\Big[\frac{1}{(p^2+m^2)}}\Big]^{alter}_{\mu,D=2}\!\!\!\!\!\!\!\!\!\!&=&\!\!\!\!\!\partial_{X}\big[X
\int^{\!\!\ ^{\mu^2}}_{1}\frac{dt}{t} \frac{1}{(X \Lambda^{2}+m^{2} t)}\big] \nonumber\\
\!\!\!\!\!\!\!\!\!\!&=&\!\!\!\!\!\frac{1}{(p^{2}+m^{2} )}-\frac{1}{(p^{2}+m^{2}\mu^2 )} 
\end{eqnarray}
This is a Pauli-Villars subtraction, however without any additionnal scalar field.It is checked 
that the final momentum integration gives the very same result as in Eq.(17).The same analysis and conclusion
hold at $D=4$.It is known \cite{ITZ} that in $ \phi^{4}_{4}$ theory the replacement of Eq.(21) makes every diagram finite
but the one loop tadpole which is treated  in Eqs(18,19).The OPVD treatment gives therefore a completely finite perturbative expansion. \\

\section{INFRARED EXTENSION OF $T(X)$} 

\subsection{Test function in the infrared.}

We consider a distribution $T(X)$ singular at the origin of ${\mathbb R}^{d}$ in the sense of the first paragraph and homogeneous,
that is $T(\frac{X}{t})=t^{(k+d)} T(X)$, where $k$ is the singular order defined in Eq.(7).The test function  which
vanishes at the origin with all its derivative can be written as $f^{<}(X)=w(X)f^{>}(X)$ with $w(X)=\chi(h-\| X \|+1,h)$.
As for the UV-case $w(\frac{X}{t})$ effectively cuts the t-integration {\it ie} $ \| X \|(\mu^2-1) \equiv \tilde{\mu}\| X \|
< t < 1$. It gives
\begin{eqnarray} 
<\!\widetilde{T}^{<},1\!\!>\!\!\!\!\!&=&\!\!\!\!\!(-)^{k+1}(k+1)\!\!\!\!\!\!{\displaystyle \sum_{\mid \beta \mid=k+1}}\!\!\!\!
\int\!\! d^{d}X  \partial^{\beta}_{X}\big[\frac{X^{\beta}}{\beta !} T(X)\nonumber \\
\!\!\!\!\!& &\!\!\!\!\!\int_{\tilde{\mu}\| X \|}^{1}\!\!\!\!\! dt \frac{(1-t)^{k}}{t}\big].
\end{eqnarray}
The $t$-integration is trivial giving  \cite{GBL03}
\begin{eqnarray}
\widetilde{T}^{<}(X)\!\!\!\!\!&=&\!\!\!\!\!(-)^{k}(k+1)\!\!\!\!\!\!{\displaystyle \sum_
{\mid \beta \mid=k+1}} \!\!\!\!\partial^{\beta}_{X}
\big[\frac{X^{\beta}}{\beta !} T(X)\log(\tilde{\mu}\| X \|)\big] \nonumber\\
\!\!\!\!\!& &\!\!\!\!\!+\frac{(-)^{k}}{k !}H_{k}{\displaystyle \sum_
{\mid \beta \mid=k}} C^{\beta} \delta^{(\beta)}(X), 
\end{eqnarray}
with $H_{k}={\displaystyle\sum_{p=1}^{k}}\frac{(-1)^{(p+1)}}{p}\left(\begin{array}{c}k\\p\end{array}\right)=\gamma+\psi(k+1)$ and
$C^{\beta}=\int_{(\| X \|=1)} T(X) X^{\beta} dS$.

\subsection{Application: massive scalar field propagator at $D=2$ from perturbative mass expansion.}

 The free massive scalar field propagator $D_{F}(x)$ is a known function.Its zero mass expression $D_{F}^{0}(x)$ is also known 
 from Conformal Field Theory (CFT).However it is also a well-known fact that any attempt to derive $D_{F}(x)$ from a perturbative
 expansion in the mass is conventionnaly hopeless because of crippling infrared divergences.Taking into account the OPVD nature
 of the field,$D_{F}(x)$ obeys the following mass expansion
\begin{eqnarray}
D_{F}(x)\!\!\!\!\!&=&\!\!\!\!\!D_{F}^{0}(x)-m^2\int \frac{d^2 p}{(2 \pi)^2}\frac{e^{i p.x}}{p^4}(f^{<}(p^2))^{4}\nonumber \\\
\!\!\!\!\!& &\!\!\!\!\!+m^4 \int \frac{d^2 p}{(2\pi)^2}\frac{e^{i p.x}}{p^6}(f^{<}(p^2))^{6}+...  
\end{eqnarray}
 From $\widetilde{T}^{<}(X)$ with $X=\frac{p^2}{\Lambda^2}$ one finds 
 \begin{eqnarray}
\widetilde{\Big[\frac{1}{(p^2)^{(k+1)}}}\Big]\!\!\!\!\!&=&\!\!\!\!\!\frac{(-)^{k}}{k
!}\frac{\partial^{k+1}}{\partial(p^2)^{k+1}}\big[\log(\frac{p^2}{\Lambda^2})\big] \nonumber \\
\!\!\!\!\!& &\!\!\!\!\!+ 2\frac{(-)^{k}}{k!}H_{k}\delta^{(k)}(p^2).
\end{eqnarray} 
The Fourier transform of this distribution writes
\begin{eqnarray}
\int \frac{d^2 p}{(2 \pi)^2}\frac{e^{i p.x}}{\widetilde{\big[(p^2)^{(k+1)}\big]}}\!\!\!\!\!&=&\!\!\!\!\!\frac{(-)^{k}}{2\pi(k
!)^2}(\frac{\mid x\mid^2}{4})^{k}\Big[\psi(k+1) \nonumber \\
\!\!\!\!\!& &\!\!\!\!\!-\log(\frac{\Lambda\mid x\mid}{2})\Big]
\end{eqnarray}
For $k=0$  this is $-\frac{1}{2\pi}\Big[\gamma+\log(\frac{\Lambda\mid x\mid}{2})
\Big]\equiv D_{F}^{0}(x) $, giving $\Lambda \equiv m$.The overall expression for $D_{F}(x)$ is then
\begin{eqnarray}
D_{F}(x)\!\!\!\!\!&=&\!\!\!\!\!\frac{1}{2\pi}{\displaystyle \sum_{k=0}^{\infty}}\frac{\Big[\psi(k+1)-\log(\frac{m \mid x\mid}{2})\Big]}{(k
!)^2}\big[\frac{m^2 \mid x\mid^2}{4}\big]^{k} \nonumber \\
 \!\!\!\!\!&=&\!\!\!\!\!\frac{1}{2\pi} K_{0}(m \mid x\mid)
\end{eqnarray}
This gratifying exact result carries over to $D=4$ as well.

\section{CONCLUSIONS.}

In Euclidean metric we have shown that treating fields as OPVD with appropriate test functions leads to  
finite actions and well defined observables free of divergences.A finite RG-analysis needs only to be 
performed with respect to the scale parameter present in the test function.In Minkowskian metric,where
the implementation of causality is the major issue,Epstein,Glaser and followers \cite{EG73,Mphys,Scharf}
have shown that Taylor substractions are equivalent to symmetry preserving dispersion relations 
with,as we have shown here,the possible interpretation in terms of Pauli-Villars type of subtractions at the level 
of propagators,but without the introduction of new fields.The link with dimensional regularization through analytic 
continuation of powers of propagators has also been established \cite{GBL03} thereby showing that all known symmetry- 
preserving regularizations are rooted in the proper OPVD treatment of fields.It has an immediate application in the 
calculation of abelian anomalies as reported in our $LC2004$ meeting \cite{GW04}.Other important features of the Bogoliubov-Epstein-Glaser
construction are the absence of overlapping phenomena in higher order contributions and the possibility to implement arbitrary symmetries via 
the quantum Noether method \cite{HUR} without the otherwise unavoidable necessity to regularize infinite contributions.
These recent developments of Epstein and Glaser's causal approach make it extremely plausible that a finite symmetry-preserving LCQFT 
could be envisaged on the basis of an iterative construction of the $S$-matrix
and a causality conditioned finite regularization using the OPVD treatments of fields advocated in this contribution.


\begin{thebibliography}{9}
\bibitem{Bolo}N.N. Bogoliubov, D.V. Shirkov,
 "Introduction to the Theory of Quantized Fields',New York,
  J. Wiley \& Sons, Publishers, Inc.,  3rd edition 1980. 
\bibitem{Schw}S. Schweber, "An Introduction to Relativistic Quantum Field Theory", New York,
 Harper \& Row, Publishers, Inc., 1961.
\bibitem{THEW}T. Heinzl, E.Werner, Z. Phys. {\bf C62} (1994) 521.  
\bibitem{GSW}
 S. Salmons, P. Grang\'e, E. Werner, Phys. Rev. {\bf D57} (1998) 4981; Phys. Rev. {\bf D60} (1999) 067701; 
 Phys. Rev. {\bf D65} (2002) 125015.
\bibitem{GUW} P. Grang\'e, P. Ullrich, E. Werner, Phys. Rev. {\bf D57} (1998) 4981. 
\bibitem{EG73} H. Epstein, V. Glaser,
   Ann. Inst. Henri Poincar\'e {\bf XIXA} (1973) 211 .   
\bibitem{ULL}P. Ullrich, J. Math. Phys.{\bf 45,8} (2004) 3109.
\bibitem{Mphys} A.N. Kuznetsov, F.V. Tkachov, V.V. Vlasov, {\bf hep-th/9612037};
 J. Prange, J. Phys. {\bf A 32.} (1999) 2225;
 M. D\"{u}tsch, K. Fredenhagen, Com. Math. Phys. {\bf 219.} (2001) 5;
 G. Pinter, Annalen Phys. {\bf (Ser. 8) 10.} (2001) 333;
 J.M. Garcia-Bonda, Math. Phys. Anal. Geom. {\bf 6.} (2003) 59; 
\bibitem{GBL03} J.M. Garcia-Bonda, S. Lazzarini, J. Math. Phys.  {\bf 44.} (2003) 3863.
\bibitem{Sch63} L. Schwartz, " Th\'eorie des Distributions", Hermann, Paris (1963). 
\bibitem{ITZ} C. Itzykson, J.B. Zuber, "Quantum Field Theory" McGraw-Hill Inc,New York (1980).
\bibitem{KOB} S. Kobayashi, K. Nomizu, "Foundations of Differential Geometry", Interscience Tracts in Pure and Applied
Mathematics, N 1, Vol. 15 (1963);\\
M. Spivak, "A Comprehensive Introduction to Differential Geometry", Waltham, Mass.  Brandeis University.
\bibitem{Scharf} G. Scharf, "Finite QED:the causal approach",Springer Verlag (1995). 
\bibitem{GW04} P. Grang\'e, E. Werner, Few-Body Systems {\bf 35} (2005) 103.
\bibitem{HUR}T. Hurth, K. Skenderis,  Nucl. Phys. {\bf B541} (1999) 566.
\end{thebibliography}
\end{document}